\documentstyle[12pt]{article}
\begin{document}
\baselineskip=14pt

\begin{center}
{\Large {\bf An Explanation on Negative Mass-Square 
of Neutrinos}}

\vskip 0.8cm
            Tsao Chang \\ 
  Center for Space Plasma and Aeronomy Research\\
  University of Alabama in Huntsville\\
            Huntsville, AL 35899\\
      Email: changt@cspar.uah.edu\\

\vskip 0.5cm
Guangjiong Ni \\
Department of Physics, Fudan University \\
 Shanghai, 200433, China \\

\vskip 0.5cm
\end{center}

\noindent
{\bf Abstract:}
It has been known for many years that the measured mass-square of
neutrino is probably negative.  For solving this puzzle, we have
further investigated the hypothesis that neutrinos are superluminal
fermions.  A new Dirac-type equation is proposed and a tachyonic
quantum theory is briefly discussed. This equation is equivalent to
two Weyl equations coupled together via nonzero mass while respecting
the maximum parity violation, and it reduces to one Weyl equation
when the neutrino mass becomes zero.

\vskip 0.5cm
\noindent
PACS number: 14.60.Lm, 14.60.Pq, 14.60.St

\newpage
\baselineskip=14pt
\noindent
{\bf 1. Introduction}

    The square of the neutrino mass is measured in tritium beta decay 
experiments by fitting the shape of the beta spectrum near endpoint. 
  In many experiments, it has been found to be negative.
  Most recent data are listed in 
"Review of Particle Physics, 2000" [1] and references therein.  The 
weighted average from two experiments reported in 1999 [2-3] is
$$	 m^2 (\nu_e) = -2.5 \pm  3.3 \,  eV^2    \eqno (1) $$	
However, other nine measurements from different experiments in 1991-1995 
are not used for averages.  For instance, a value of $m^2$ ($\nu_e$) = 
-130 $\pm$ 20 $eV^2$ with 95$\%$ confidence level was 
measured in LLNL in 1995[4].

    Furthermore, the pion decay experiment also 
obtained a negative value for $\mu$-neutrinos [5].
$$  m^2(\nu_{\mu}) = - 0.016 \pm 0.023 \, MeV^2  \eqno (2)   $$	

   The negative value of the neutrino mass-square simply means:
     $$  E^2/c^2 - p^2 = m^2 (\nu_e)c^2 < 0	\eqno  (3)  $$      
The right-hand side in Eq. (3) can be rewritten as (- $m_s^2 c^2$), 
then $m_s$ has a positive value.  Eq. (1) and (2) suggests that 
 neutrinos might be particles faster than light, no matter how small
the $m_s$ is.  This possibility is further investigated in this paper.  

    Based on special relativity and known as re-interpretation rule, 
superluminal particles were proposed by Bilaniuk et al in the Sixties
[6-8].  The sign of 4-D world line element, $ds^2$, is associated with 
three classes of particles.   For simplicity, let $dy = dz = 0$, then
$$ \hskip 4.2cm	 > 0 \hskip 0.5 cm  \rm{Class I \,(subluminal 
   \,particles)} $$
$$  ds^2 =c^2dt^2 - dx^2 = 0 \hskip 0.5cm  \rm{Class II 
  \, (photon)}   \eqno (4) $$
$$  \hskip 4.5 cm < 0    \hskip 0.5 cm  
   \rm{Class III \,(superluminal \, particles)}    $$
  For Class III particles, i.e. superluminal particle, the relation of 
momentum and energy is shown in Eq.(3). The negative value on the 
right-hand side of Eq. (3) for superluminal particles means that 
$p^2$ is greater than $(E/c)^2$.  The velocity of a 
superluminal particle, $u_s$, is greater than the speed of light.  The 
momentum and energy in terms of $u_s$ are as follows:
    $$	p =  \frac{m_s u_s}{\sqrt{u_s ^2 /c ^2  - 1}},  \quad 		
    	E = \frac{m_s c^2}{\sqrt{u_s ^2 /c ^2  - 1}}     \eqno  (5) 
$$	
where the subscript $s$ means superluminal particle, i.e. tachyon.  
From Eq. (5), it is easily seen that when $u_s$ is increased, 
both of $p$ and $E$ would be decreased.  This property is opposite 
to Class I particle.

 Any physical reference system is built by Class I particles (atoms, 
molecules etc.), which requires that any reference frame must move 
slower than light.  On the other hand, once a superluminal particle 
is created in an interaction, its speed is always greater than 
the speed of light.  Neutrino is the most possible candidate for a 
superluminal particle because it has left-handed spin in any reference 
frame.  On the other hand, anti-neutrino always has right-handed spin.

The first step in this direction is usually to introduce an 
imaginary mass, but these efforts could not reach a point for 
constructing a consistent quantum theory.  Some early investigations
of a Dirac-type equation for tachyonic fermions can be found in 
Ref.[9].  An alternative approach was investigated by Chodos et al.
[10].  They examined the possibility that neutrinos might be
tachyonic fermions.  A form of the lagrangian density for tachyonic 
neutrinos was proposed.  Although they did not obtain a satisfatory 
quantum theory for tachyonic fermions, they suggested that more 
theoretical work would be needed to determine physically acceptable 
tachyonic theory.\\

\noindent
{\bf 2. A new Dirac-type equation}

    In this paper, we will start with a different approach to derive a 
new Dirac-type equation for tachyonic neutrinos.  In order to 
avoid introducing imaginary mass, Eq. (3) can be rewritten as
    $$ E = (c^2p^2 -  m_s^2c^4 )^{1/2}  \eqno  (6)  $$
where $m_s$  is called proper mass,  for instance, $m_s(\nu_e)$ = 1.6 
eV  from Eq. (1).  We follow Dirac's search [11], Hamiltonian 
must be first order in momentum operator ${\hat p}$ :
  $$  \hat E = - c ({\vec \alpha} \cdot {\hat p}) + \beta_s m_s c^2 	
     \eqno (7)  $$
with  ($\hat E = i\hbar \partial /\partial t , {\hat p} = 
-i \hbar \nabla $), where ${\vec \alpha} = (\alpha_1, \alpha_2, \alpha_3 
)$ and $\beta_s$ are 4$\times$ 4 matrix, which are defined as\\
  $$ {\alpha_i} = \left(\matrix{0&\sigma_i\cr
                         \sigma_i&0\cr}\right),  \quad
   \beta_s = \left(\matrix{0&I\cr
                         -I&0\cr}\right)  \eqno (8)   $$
where $\sigma_i$ is 2$\times$2 Pauli matrix, $I$ is 2$\times$2 unit 
matrix.  Notice that $\beta_s$ is a new matrix, which 
is different from the one in the traditional Dirac equation.  
the relation between the matrix $\beta_s$ and the traditional matrix 
$\beta$ is as follows:
$$  \beta_s = \beta \gamma_5     \eqno (9)  $$
where
$$   {\beta} = \left(\matrix{I&0 \cr
     0&-I \cr}\right), \quad   
{\gamma_5} = \left(\matrix{0&1 \cr
     1&0 \cr}\right)     \eqno	(9a)   $$ 

 When we take square for both sides in Eq.(7), and consider the 
following relations:
 $$ \alpha_i \alpha_j + \alpha_j \alpha_i = 2 \delta_{ij}    $$
      $$	\alpha_i \beta_s + \beta_s \alpha_i =  0  $$
      $$       \beta_s^2 = -1	\eqno    (10)		$$
the relation in Eq. (3) or Eq. (6) is reproduced.  Since Eq. (6) is 
related to Eq. (5), this means $\beta_s$ is a 
right choice to describe neutrinos as superluminal particles. 

Denote the wave function as
  $$ \Psi = \left(\matrix{\varphi ({\vec x},t)\cr
                         \chi ({\vec x},t)\cr}\right) \quad
with \quad
   \varphi = \left(\matrix{\varphi_1\cr
                         \varphi_2\cr}\right), \quad
    \chi = \left(\matrix{\chi_1\cr
                         \chi_2\cr}\right)  \eqno (11)  $$
From Eq. (7), the complete form of the new Dirac-type equation 
becomes
$$  \hat E \Psi = -c({\vec \alpha} \cdot {\hat p})\Psi + 
           \beta_s m_s c^2 \Psi 	     \eqno (7a)  $$
Since $\beta_s^2 = -1$ in the Eq.(7a), the new Dirac-type 
equation is different from the traditional Dirac equation in any 
covariant representation in terms of the $\gamma$ matrices.

We now study the spin-1/2 property of neutrino as a tachyonic 
Fermion.  Eq. (7a), can be rewritten as a pair of two-component 
equations:
$$ i\hbar \frac{\partial \varphi}{\partial t} = ic 
    \hbar {\vec \sigma} \cdot  \nabla \chi + m_s c^2 \chi   $$
	$$ i\hbar \frac{\partial \chi}{\partial t} = ic \hbar 
\vec{\sigma} \cdot \nabla \varphi - m_s c^2 \varphi  
\eqno (12)   $$
Eq. (12) is an invariant under the space-time inversion transformation 
with (${\vec x} \rightarrow -{\vec x}, t \rightarrow -t $) [12,13],
$$ \varphi (-{\vec x}, - t) \rightarrow \chi ({\vec x}, t), 
\quad  \chi (-{\vec x}, -t ) \rightarrow \varphi ({\vec x}, t) 
\eqno (13)   $$
It means that Eq.(12) obeys CPT theorem.  From the equation (12), 
the continuity equation can be derived:
	$$  \frac{\partial \rho}{\partial t} +
           \nabla \cdot {\vec j} = 0  \eqno (14)   $$
and we have
  $$ \rho = \varphi^{\dag} \chi + \chi^{\dag} \varphi ,  \quad
  {\vec j} = -c(\varphi^{\dag} {\vec \sigma} \varphi + \chi^{\dag} 
 {\vec \sigma} \chi)    \eqno (15)  $$
where $\rho$ and $\vec j$ are the probability density and current; 
$\varphi^{\dag}$ and $\chi^{\dag}$ are the Hermitian adjoint 
of $\varphi$ and $\chi$ respectively.

Eq.(15) can be rewritten as
  $$ \rho = \Psi^{\dag}  \gamma_5 \Psi ,  \quad
  {\vec j} = c(\Psi^{\dag}  \gamma_5 {\vec \alpha} \Psi) 
     \eqno (15a)  $$
It is easy to see that the probability density $\rho$ is positive 
definite when the components in $\varphi$ and $\chi$ are positive.

Considering a plane wave along the $z$ axis for a left-handed particle 
$({\vec \sigma} \cdot {\vec p})/p = -1$, the equations (12) 
yields the following solution:
  $$	 \chi = \frac{cp - m_s c^2}{E} \varphi   \eqno	(16) $$
    We now consider a linear combination of $\varphi$ and $\chi$, 
  $$   \xi= {1 \over {\sqrt 2}} (\varphi + \chi)  , \quad
      \eta = {1 \over {\sqrt 2}} (\varphi - \chi)  \eqno (17)  $$
where $\xi ({\vec x},t)$ and $\eta ({\vec x},t)$ are two-component 
spinor functions. In terms of $\xi$ and $\eta$, Eq. (15) becomes 
  $$ \rho = \xi^{\dag} \xi - \eta^{\dag} \eta ,  \quad
   {\vec j} = -c(\xi^{\dag} {\vec\sigma} \xi + \eta^{\dag} 
{\vec \sigma} \eta)     \eqno (18)  $$
In terms of Eq. (17), the equation (12) can be rewritten in the
Weyl representation: 
$$ i\hbar \frac{\partial \xi}{\partial t} = ic \hbar {\vec \sigma}
    \cdot \nabla \xi - m_s c^2 \eta   $$
$$ i\hbar \frac{\partial \eta}{\partial t} = -ic \hbar {\vec \sigma}
     \cdot  \nabla \eta + m_s c^2 \xi  \eqno (19)   $$
In the above equations, both $\xi$ and $\eta$ are coupled via nonzero 
$m_s$.

    In order to compare Eq. (19) with the well known two-component Weyl 
equation, we take a limit $m_s = 0$, then the first equation in Eq.(19)
 reduces to
	$$   \frac{\partial \xi_\nu}{\partial t} = c{\vec \sigma} \cdot
           \nabla \xi_\nu      \eqno (20)   $$
and the second equation in Eq. (19) vanishes because $\varphi = \chi$
for the massless limit.

 Eq. (20) is the two-component Weyl equation for describing neutrinos, 
which is related to the maximum parity violation discovered in 1956 
by Lee and Yang [14,15].  They pointed out that no experiment had 
shown parity to be good symmetry for weak interaction.  Now we see 
that, in terms of Eq.(19), once if neutrino has some mass, no 
matter how small it is, two equations are coupled together via 
the mass term while still respecting the maximum parity violation. 

\vskip 0.6cm
\noindent
{\bf 3. Remarks}

In this paper, we have further investigated the hypothesis 
that neutrinos are tachyonic fermions.  A new Dirac-type equation is 
proposed and a tachyonic quantum theory is briefly discussed, which 
can be used to solve the puzzle of negative mass-square of neutrinos.

    Notice that the matrix $\beta_s$ in the new Dirac-type equation  
is not a $4 \times 4$ hermitian matrix.  However, based 
on the above study, we now realize that the violation of hermitian 
property is related to the violation of parity.  Though a non-
hermitian Hamiltoian is not allowed for a subluminal particle, it does 
work for superluminal neutrinos. 

   In the light of the above theoretical framework, the consequences 
derived from the new Dirac-type equation agree with all known 
properties of neutrinos.  Therefore, we have reached a conclusion 
that neutrinos are tachyonic fermions with permanent helicity if the 
neutrino mass-square is negative.  Therefore, more accurate tritium 
beta decay experiments are needed to further determine the neutrino 
mass-square. 

  According to special relativity [16], if there is a superluminal 
particle, it might travel backward in time.  However, a re-interpretation
rule has been introduced since the Sixties [6-8].  Another approach is 
to introduce a kinematic time under a non-standard form of the Lorentz 
transformation [17-20].  Therefore, special relativity can 
be extended to space-like region, and superluminal particles are 
allowed without causality violation.

\baselineskip=14pt
\vskip 0.6cm
  We wish to thank S.Y. Zhu and Y. Takahashi for helpful discussions

\end{document}